\documentclass[prb]{revtex4}% Physical Review B

\usepackage{graphicx}% Include figure files
\usepackage{dcolumn}% Align table columns on decimal point
\usepackage{bm}% bold math

\begin{document}
\pagestyle{empty}

%\preprint{APS/123-QED}

%\title{Verification of $z$-Scaling at RHIC and Tevatron}
\title{ $z$-Scaling and Prompt $J/\psi$ Production at High-$p_T$ in $\bar pp$ at Tevatron}

\author{M.V. Tokarev\footnote{e-mail: tokarev@sunhe.jinr.ru}}
\affiliation{Joint Institute for Nuclear Research, Dubna, Moscow region, 141980, Russia
}

\begin{abstract}
New data on high-$p_T$ spectra of $J/\psi$
measured  by the CDF Collaboration at the  Tevatron are
analyzed in the framework of $z$-scaling.
$F$-dependence of $z$-presentation of the data is used to construct
the scaling function of $J/\psi$.
The energy independence of the scaling function
is used to predict inclusive cross section of prompt $J/\psi$ production
in $\bar pp$ collisions at high-$p_T$.
The concept of $z$-scaling is considered to reflect the general features
of high-$p_T$ particle production in hadron-hadron and hadron-nucleus collisions.
Violation of $z$-scaling is suggested to consider as a signature of new physics.
\end{abstract}

%\pacs{Valid PACS appear here}
%Use showkeys class option if keyword
\maketitle

%\section{\label{sec:level1}Introduction}

\begin{center}
{1. INTRODUCTION}
\end{center}

%\protect}
%\textbackslash
%\textbackslash}
Primary goal of all new accelerators is to discover new physics
phenomena. They can give impulse for development
of physics theory and extend our understanding of micro world.
The origin of particle structure at small scales,
distributions of mass, charge and spin in space-time,
features of particle-to-particle transitions,  are of the most
enigmatic and least understandable physics phenomena.
Therefore, search for scaling regularities in high energy particle
collisions is a subject of intense investigations
\cite{Feynman,Bjorken,Bosted,Benecke,Baldin,
Stavinsky,Leksin,KNO,Matveev,Brodsky}.

Properties of particle formation are believed to reveal
themselves more clearly at high energy $\sqrt s $ and transverse momenta $p_T$.
It is considered also that partons
produced in hard scattering retain information about primary
collision during  hadronization.
It is known that the mechanism
of particle formation is modified
by nuclear medium  and can be sensitive to the phase transition
of nuclear matter. Therefore, the  features of high-$p_T$
single inclusive particle spectra of hadron--hadron and  hadron--nucleus
collisions  are of interest to search for new physics phenomena
in elementary processes (quark compositeness, fractal space-time,
extra dimensions),  signatures  of exotic state of nuclear matter
(phase transitions, quark-gluon plasma), complementary restrictions
for theory (nonperturbative QCD effects, physics beyond Standard Model).

New feature ($z$-scaling) of high-$p_T$ particle production in
hadron-hadron and hadron-nucleus collisions at high energies was
established in  \cite{Zscal,zppg}.
 The scaling function $\psi$ and scaling variable $z$ are expressed via
experimental quantities such as the inclusive cross section
$Ed^3\sigma/dp^3$ and the multiplicity density of charged
particles $dN/d\eta$. The $z$-presentation of data is found to reveal
symmetry properties (energy independence, A- and F-dependence, power
law).
%The properties of $\psi$ at high $z$ are assumed  to be
%relevant to the structure of space-time at small scales \cite{Zbor,Nottale}.
The function $\psi(z)$ is interpreted as the
probability density to produce a particle with a formation length
$z$. The concept of $z$-scaling and the method of data analysis
are developed for description  of different particles (charged
%\cite{Z99,Z01,Toivonen}
and neutral
% \cite{Rog1,Rog2}
 hadrons, direct photons
% \cite{Potreben,Efimov},
jets
 %  \cite{Dedovich}
 )
 produced in high energy hadron-hadron and hadron-nucleus interactions.
 The proposed method is complementary to a method of direct
calculations developed in the framework of QCD
 %  \cite{QCD}
and methods based on the Monte Carlo generators
 % \cite{JETRAD,PYTHIA,HERWIG,ISAJET,DPNJET,FRITIOF,HIJING,NEXUS}
 .
Therefore we consider that the use of the method allow to reduce
some theoretical uncertainties which are ambiguously estimated by theory.

Recently new data \cite{Abe2} on cross section of prompt $J/\psi$ production
in  $\bar pp$ collisions in the central rapidity range $|\eta_{J/\psi}|<0.6$
 at $\sqrt s =1.96$~TeV are measured by the CDF Collaboration.
The data and the other one \cite{Abe1} obtained in Run I give
new possibility to verify  models of $J/\psi$ production mechanisms.
In the color singlet model \cite{Baier} the charmoniunm meson retains
the quantum numbers of the produced $\bar cc$ pair and each $J/\psi$ state
is directly produced in hard scattering of color singlet
via the corresponding subprocess. As shown in \cite{Abe1} the model predictions
are in a disagreement with experimental data on $J/\psi$ and $\psi(2S)$
production rates. A color octet model \cite{Octet1,Octet2} was suggested
to explain the discrepancy.
The color octet mechanism takes into  account the production of $\bar cc$ pair
in a color octet configuration. The color octet state evolves  into a color singlet
state via emission of a soft gluon. The color octet contributions were found in \cite{Octet3}
from a fit to CDF data \cite{Abe1}.

We use a complementary method of data analysis
and  present the results of analysis of CDF data \cite{Abe2} on prompt $J/\psi$
production  in $\bar pp$ collisions in Run I and Run II at
Tevatron in the framework of $z$-scaling.  Properties of
$z$-scaling, the energy independence and $F$-dependence, are used
to construct the scaling function for $J/\psi$. The inclusive
cross sections of $J/\psi$ production in $\bar pp$ collisions at high-$p_T$
are estimated. The results can be useful to test phenomenological models
of $J/\psi$ production mechanism and obtain new restrictions on a gluon distribution function.
The possibility to search for the scaling violation  at the Tevatron is suggested as well.

\vskip 15mm
\begin{center}
{2. $Z$-SCALING}
\end{center}

In the section underlying ideas of $z$-scaling,
a general scheme of data $z$-presentation  and physical
meaning of the scaling function $\psi(z)$
and the scaling variable $z$ are discussed.

\begin{center}
{2.1. Basic principles: locality, self-similarity, fractality}
\end{center}

The idea of $z$-scaling is based on the assumptions \cite{Stavinsky}
that gross feature of inclusive particle distribution of the
process (\ref{eq:r1}) at high energies can be described in terms
of the corresponding kinematic characteristics
\begin{equation}
M_{1}+M_{2} \rightarrow m_1 + X
\label{eq:r1}
\end{equation}
of the constituent subprocess  written in the symbolic form (\ref{eq:r2})
\begin{equation}
(x_{1}M_{1}) + (x_{2}M_{2}) \rightarrow m_{1} +
(x_{1}M_{1}+x_{2}M_{2} + m_{2})
\label{eq:r2}
\end{equation}
satisfying the condition
\begin{equation}
(x_1P_1+x_2P_2-p)^2 =(x_1M_1+x_2M_2+m_2)^2.
\label{eq:r3}
\end{equation}
The equation is the expression of locality of  hadron interaction at
constituent level. The $x_1$ and $x_2$ are fractions of the incoming
momenta $P_1$ and $P_2$  of  the colliding objects with the masses $M_1$
and $M_2$. They determine the minimum energy, which
is necessary for production of the secondary particle with
the mass $m_1$ and the four-momentum $p$.
The parameter $m_2$ is introduced to satisfy the internal
conservation laws (for baryon number, isospin, strangeness, and so on).

The equation (\ref{eq:r3}) reflects minimum recoil mass hypothesis in the
elementary subprocess.
To connect kinematic and structural
characteristics of the interaction, the quantity
$\Omega$ is introduced. It is chosen in the form
\begin{equation}
\Omega(x_1,x_2) = m(1-x_{1})^{\delta_1}(1-x_{2})^{\delta_2},
\label{eq:r5}
\end{equation}
where $m$ is a mass constant and $\delta_1$ and $\delta_2$
are factors relating to the anomalous fractal dimensions  of
the colliding objects. The fractions $x_{1}$ and
$x_{2}$  are determined  to maximize the value of $\Omega(x_1,x_2)$,
simultaneously fulfilling the condition (\ref{eq:r3})
\begin{equation}
{d\Omega(x_1,x_2)/ dx_1}|_{x_2=x_2(x_1)} = 0.
\label{eq:r6}
\end{equation}
The fractions  $x_{1}$  and $x_2$ are equal to unity along
the phase space limit and
cover the full phase space accessible at any energy.

Self-similarity is a scale-invariant property connected with dropping of
certain dimensional quantities out of physical picture of the interactions.
It means that dimensionless quantities for the description of physical processes
are  used. The scaling function
$\psi(z)$ depends in a self-similar manner on the single dimensionless variable
$z$. It is expressed  via the invariant cross section
$Ed^3\sigma/dp^3$ as follows

\begin{equation}
\psi(z) = -{ { \pi s} \over { (dN/d\eta) \sigma_{in}} } J^{-1} E { {d^3\sigma} \over {dp^3}  }
\label{eq:r7}
\end{equation}
Here, $s$ is the center-of-mass collision energy squared, $\sigma_{in}$ is the
inelastic cross section, $J$ is the corresponding Jacobian.
The factor $J$ is the known function of the
kinematic variables, the momenta and masses of the colliding and produced particles.

The function $\psi(z)$ is normalized as follows
\begin{equation}
\int_{0}^{\infty} \psi(z) dz = 1.
\label{eq:r8}
\end{equation}
The relation allows us to interpret the function $\psi(z)$
as a probability density to produce
a particle with the corresponding value of the variable $z$.
We would like to emphasize that  the existence of the function  $\psi(z)$
depending on a single dimensionless  variable $z$ and
revealing scaling properties is not evident in advance.
Therefore the method proposed for data analysis in $z$-presentation
could be only proved  a posteriori.

Principle of fractality states that variables used in the
description of the process diverge in terms of the resolution.
This property is characteristic for the scaling variable
\begin{equation}
z = z_0 \Omega^{-1},
\label{eq:r9}
\end{equation}
where
\begin{equation}
z_0 = \sqrt{ \hat s_{\bot}} / (dN/d\eta).
\label{eq:r10}
\end{equation}
The variable $z$ has character of a fractal measure.
For the given production process (\ref{eq:r1}),
its finite part $z_0$ is the ratio
of the transverse energy released in the
binary collision of constituents (\ref{eq:r2})
and the average multiplicity density $dN/d\eta|_{\eta=0}$.
The divergent part
$\Omega^{-1}$ describes the resolution at which the collision of
the constituents can be singled out of this process.
The $\Omega(x_1,x_2)$ represents relative number of all initial
configurations containing the constituents which carry fractions
$x_1$ and $x_2$ of the incoming momenta.
The $\delta_1$ and $\delta_2$ are the anomalous fractal
dimensions of the colliding objects (hadrons or nuclei).
The momentum fractions $x_1$ and $x_2$ are determined in a way to
minimize the resolution $\Omega^{-1}(x_1,x_2)$ of the fractal
measure $z$ with respect to all possible sub-processes
(\ref{eq:r2}) subjected to the condition (\ref{eq:r3}).
The variable $z$ was interpreted  as a particle formation length.

 The scaling function of high-$p_T$ particle production, as  shown below,
 is described  by the power law, $\psi(z) \sim z^{-\beta} $.
 Both quantities, $\psi$ and $z$, are scale dependent.
 Therefore we consider the high energy   hadron-hadron interactions
 as interactions of fractals. In the
 asymptotic  region the internal structure of particles, interactions of their constituents and
 mechanism of real particle formation manifest self-similarity and fractality
 over a wide scale range.

\vskip 10mm
\begin{center}
{2.2. Prompt $J/\psi$ production in $\bar pp$}
\end{center}

In this section we analyze the data \cite{Abe1,Abe2}
on inclusive cross section of prompt $J/\psi$ production in $\bar pp$ collisions.
Firstly the experimental data \cite{Banner} on inclusive cross section
of $\pi^0$-mesons produced in $\bar pp$
collisions obtained  by the UA1 Collaboration  at $\sqrt s =540$~GeV
are used to construct $z$-presentation of the data.
The scaling function of $\pi^0$ is shown in Fig.1(a). The dashed line
is the fit of the data. As seen from Fig.1(a) the scaling function
is described by the power law, $\psi(z)\sim z^{-\beta}$,  on the
log-log scale. The value of the slope parameter $\beta$ is found to be $5.77\pm 0.02$
over a wide $p_T$ range.
Secondly  the $F$-dependence of $z$-scaling is used to compare the
scaling function of $\pi^0$ and $J/\psi$.  We assume that scaling functions
of produced particles with different flavor content has the same asymptotic
behavior \cite{Zscal,zppg}.
The transformation of variable $z$ and scaling function $\psi$ in the form
\begin{equation}
z\rightarrow \alpha_F\cdot z, \ \ \  \psi \rightarrow {\alpha_F}^{-1} \cdot \psi
\label{eq:r11}
\end{equation}
was used. The transformation parameter $\alpha_F $ is found to be 2.25.
As seen from Fig.1(a) the scaling function of $J/\psi$  at $z>10$
reveals the asymptotic behavior such as the similar one for $\pi^0$.
We would like to note that the CDF Run I and Run II
data demonstrate a good matching  in the overlapping region in $z$-presentation.
The power law for the scaling function is observed  up  to $z \simeq
40$. Therefore the kinematic range of more preferable for
search for the scaling  violation is found to be for the range $p_T>30 $~GeV/c
at $\sqrt s = 1960$~GeV.

The energy independence of the constructed scaling function  is
used to calculate the dependence of inclusive cross section  $Ed^3\sigma/dp^3$ of
$J/\psi$ production in $\bar pp$ collisions on the transverse
momentum $p_T$ in the central rapidity range, $\theta\simeq 90^0$,
at the energy $\sqrt s =63, 200, 630, 1800$ and 1960~GeV.
The calculated results are shown by the dashed lines in Fig.1(b).
Points  ($ \star, +$ ) shown in Fig.1(b) are the corresponding
CDF Run I  and Run II data.  As seen from Fig.1(b) the strong dependence
of inclusive cross section on $\sqrt s $ increases with $p_T$.

\vskip 10mm
\begin{center}
{3. $z-p_T$ PLOT}
\end{center}

The scaling function $\psi(z)$ demonstrates
the universality over a wide kinematic range.
It describes two regimes of particle formation observed at higher energies.
The first one reveals exponential behavior at low $z$
and the second one is described by the power law,
$\psi(z) \sim z^{-\beta }$, at high $z$.

Asymptotic behavior of the scaling function is described
by the slope parameter $\beta$, which depends on the anomalous
fractal dimensions $\delta_{1,2}$ of colliding objects.
We assume that new physics effects at small scales (high-$z$) should be characterized
by the change of the fractal dimension  ("$\delta$-jump")  and consequently the other
value of the slope parameter $\beta$ at high $z$.
To determine the kinematic range which is of more
preferable  for experimental searching for $z$-scaling violation
the $z-p_T$ plot can be used.
Figure 2 shows the $z-p_T$ plot, the dependence of the variable $z$ on the
transverse momentum  $p_T$  of $J/\psi$-mesons produced in $\bar pp$ collisions
at $\sqrt s=63-14000$~GeV  and $\theta_{cms}=90^0$.
As seen from Fig.1(a) the asymptotic value of the scaling function $\psi(z)$ at $z=30$
can be reached at different values of transverse momentum $p_T$
which is corresponding to different values of collision energy $\sqrt s$.
The measurement of $J/\psi$ inclusive cross section at Tevatron
energies over the range $p_T > 20$~GeV/c could give information
on asymptotic behavior of $\psi(z)$, verify the predictions
and establish features of particle formation mechanism at high-$p_T$.

\vskip 10mm
\begin{center}
{4.  CONCLUSIONS}
\end{center}

 Experimental data on inclusive cross section for prompt
 $J/\psi$ production  in $\bar pp$  collisions  at $\sqrt s =1800$ and 1960~GeV
 obtained at Run I and Run II  with the CDF detector at Fermilab
 were analyzed in the framework of $z$-presentation.
 Using the properties of $z$-scaling, the energy independence
 and $F$-dependence of $z$-presentations, the scaling  function of
 $J/\psi$ constructed over a wide $z$-range.

 In $z$-presentation the scaling function $\psi(z)$ and scaling variable $z$
 are expressed via the experimental quantities, the invariant
 inclusive cross section  $Ed^3\sigma/dp^3$ and the multiplicity
 density of charged particles $\rho(s,\eta)$.

 The physical interpretation of the scaling function $\psi$ as a
 probability density to produce a particle with the formation length $z$
 is given. The quantity  $z$ has the property of the fractal
 measure, and $\delta_{1,2} $ is the fractal dimension describing the
 intrinsic structure of the interacting constituents revealed
 at high energies.

 The prompt $J/\psi $ spectra over a wide $p_T$ and $\sqrt s $
 range are predicted.
 The kinematic regions preferable to search for $z$-scaling violation
 in $\bar pp$ collisions at $\sqrt s = 1960$~ GeV are established.

 Thus the results of our analysis confirm the concept of $z$-scaling,
 the  possibility of its applicability for analysis of experimental
 high-$p_T$ data and search for new  physics phenomena in Run II at the Tevatron.

\vskip 5mm

\acknowledgments
 The author would like to thank Yu.Gotra for useful and stimulating  discussions
of CDF experimental data and obtained results.

\begin{figure}[t]
\begin{center}
\includegraphics[width=3.in]{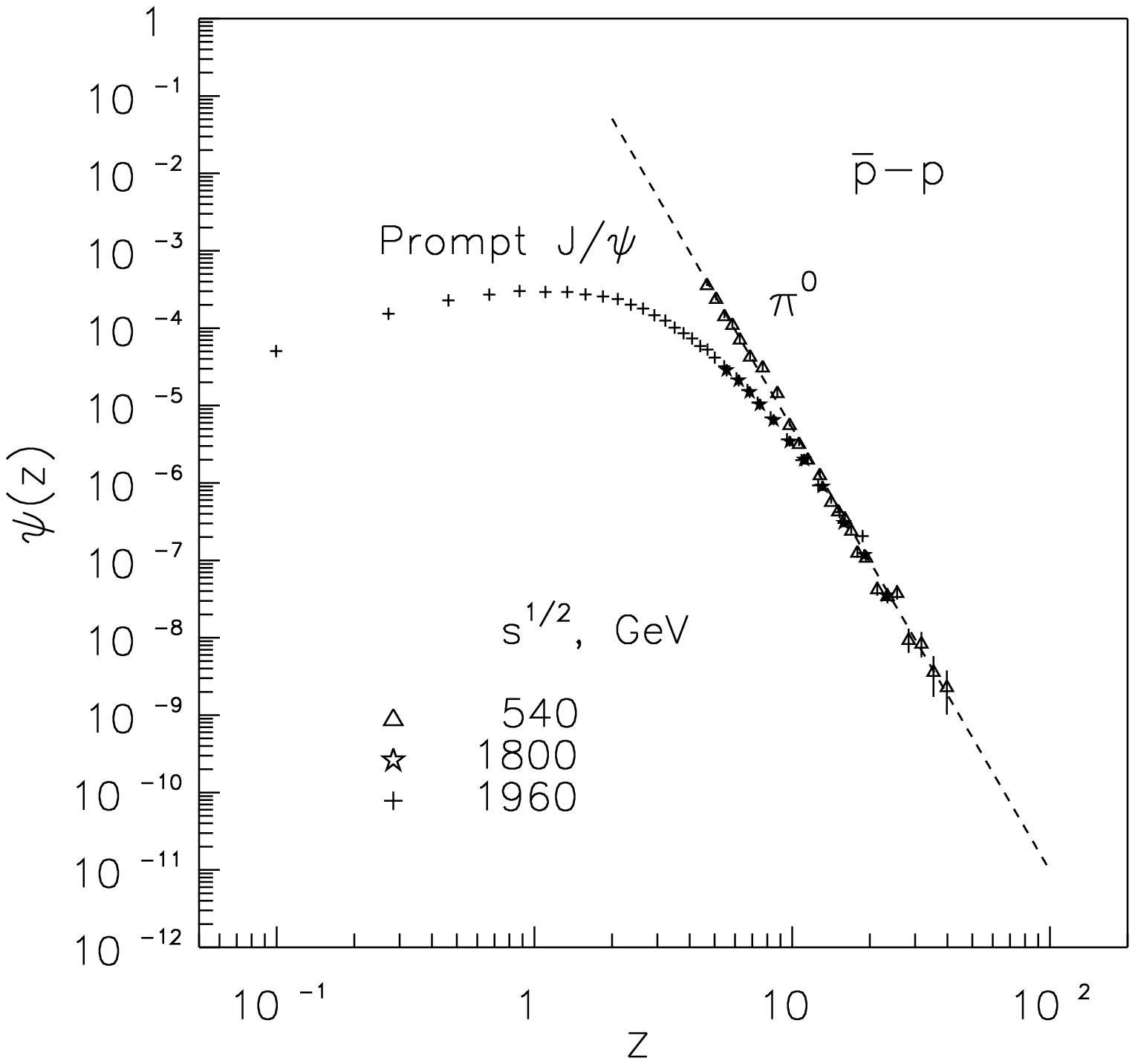}
\hspace*{1.cm}
\includegraphics[width=3.in]{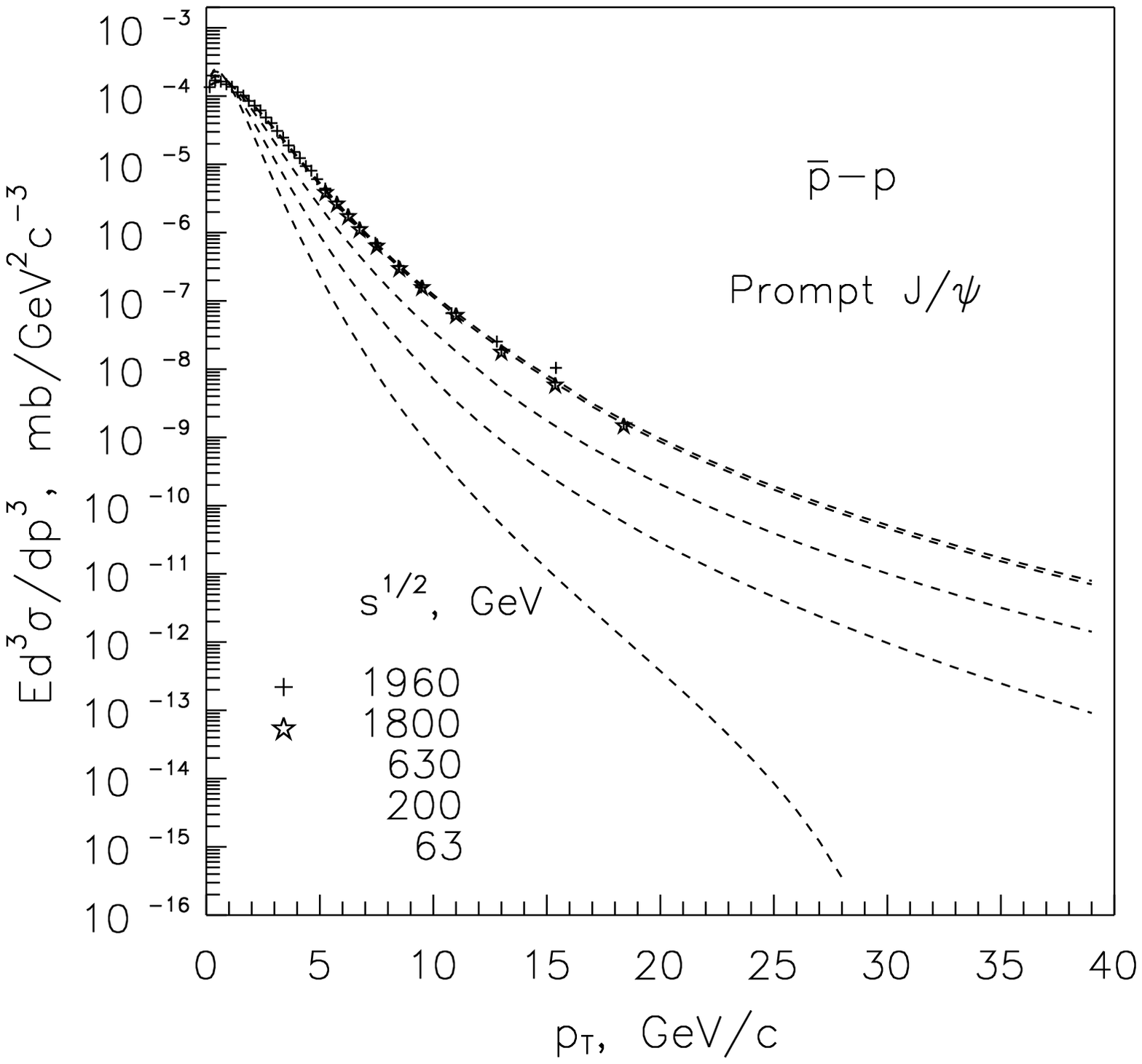}
\vskip 1mm \hspace*{1.cm} a) \hspace*{8.cm} b)\\
\end{center}
\caption{%
 (a) The scaling function of $\pi^0$- and prompt $J/\psi$-mesons produced
 in $\bar pp$ collisions in the central rapidity range
 at $\sqrt s = 540$ and 1800, 1960~GeV, respectively.
 Experimental data on inclusive cross section are taken from
\cite{Banner,Abe1,Abe2}.
(b) The inclusive differential cross sections of prompt $J/\psi$-mesons produced
 in $\bar pp$ collisions at $\sqrt s = 63, 200, 630, 1800$ and 1960~GeV
 in the central rapidity range as a functions of the transverse
momentum  $p_{T}$. Points are experimental data on inclusive cross
section taken from \cite{Abe1,Abe2}. Dashed lines are predicted results based on
$z$-scaling.
 }
\label{fig:1}
%\end{figure}
\vskip 2cm

%  Fig. 2
%\begin{figure}[c]
\begin{center}
\includegraphics[width=3.0in]{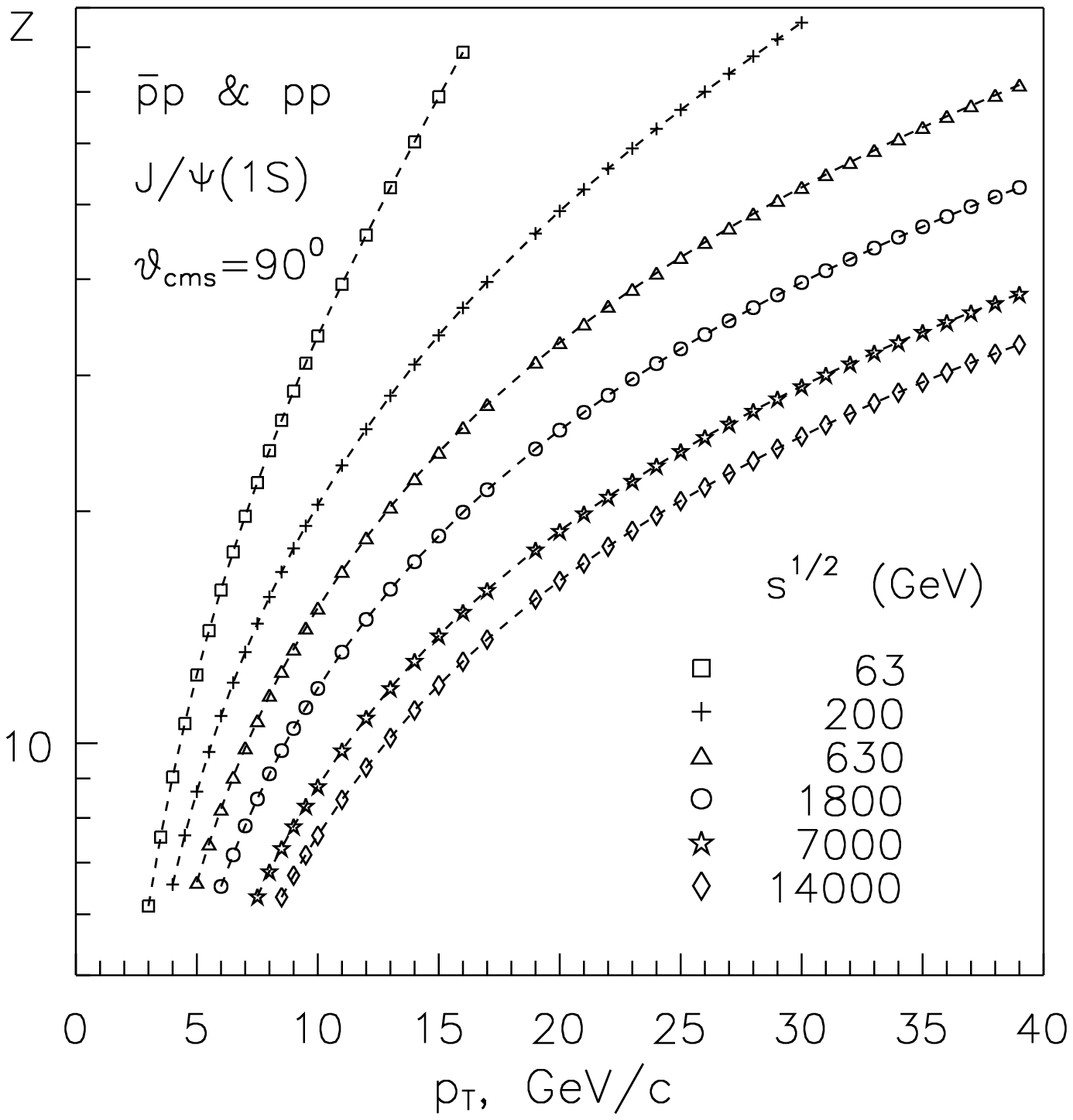}
\hspace*{1.5cm}
%\includegraphics[width=2.0in]{hgotra3.eps}
%\vskip 1mm \hspace*{0.cm} a) \hspace*{6.cm} b)\\
\end{center}
\caption{%
The dependence of the scaling variable  $z$ on the transverse momentum $p_T$
of $J/\psi$ production in $\bar pp$ and $ pp$ collisions at $\sqrt s = 63-14000$~GeV
and the angle $\theta_{cms}= 90^0$.
}
\label{fig:2}
\end{figure}

\end{document}